\documentclass[11pt]{article}
\usepackage{amsmath,amssymb}

\makeatletter
\@addtoreset{equation}{section}
\makeatother

\def\ben{\begin{equation}}
\def\een{\end{equation}}

  \let\n=\nu  \let\p=\pi

\let\C=\Chi

\def\nn{\nonumber} \def\bd{\begin{document}} \def\ed{\end{document}}
\def\ds{\documentstyle} \let\fr=\frac \let\bl=\bigl \let\br=\bigr
\let\Br=\Bigr \let\Bl=\Bigl
\let\bm=\bibitem
\let\na=\nabla
\let\pa=\partial \let\ov=\overline
\newcommand{\be}{\begin{equation}}
\newcommand{\ee}{\end{equation}}
\def\ba{\begin{array}}
\def\ea{\end{array}}
\def\ft#1#2{{\textstyle{\frac{\scriptstyle #1}{\scriptstyle #2} } }}
\def\fft#1#2{{\frac{#1}{#2}}}
\def\del{\partial}
\def\vp{\varphi}
\def\sst#1{{\scriptscriptstyle #1}}
\def\oneone{\rlap 1\mkern4mu{\rm l}}
\def\td{\tilde}
\def\wtd{\widetilde}
\def\ie{{\it i.e.\ }}
\def\dalemb#1#2{{\vbox{\hrule height .#2pt
        \hbox{\vrule width.#2pt height#1pt \kern#1pt
                \vrule width.#2pt}
        \hrule height.#2pt}}}
\def\square{\mathord{\dalemb{6.8}{7}\hbox{\hskip1pt}}}
\newcommand{\ho}[1]{$\, ^{#1}$}
\newcommand{\hoch}[1]{$\, ^{#1}$}
\newcommand{\bea}{\begin{eqnarray}}
\newcommand{\eea}{\end{eqnarray}}
\newcommand{\ra}{\rightarrow}
\newcommand{\lra}{\longrightarrow}
\newcommand{\Lra}{\Leftrightarrow}
\newcommand{\bp}{\tilde \beta^\prime}
\newcommand{\tr}{{\rm tr} }
\newcommand{\Tr}{{\rm Tr} }
\def\0{{\sst{(0)}}}
\def\1{{\sst{(1)}}}
\def\2{{\sst{(2)}}}
\def\3{{\sst{(3)}}}
\def\4{{\sst{(4)}}}
\def\5{{\sst{(5)}}}
\def\6{{\sst{(6)}}}
\def\7{{\sst{(7)}}}
\def\8{{\sst{(8)}}}
\def\n{{\sst{(n)}}}
\def\cA{{{\cal A}}}
\def\cB{{{\cal B}}}
\def\cF{{{\cal F}}}
\def\cH{{{\cal H}}}
\def\tV{\widetilde V}
\def\tW{\widetilde W}
\def\tH{\widetilde H}
\def\tE{\widetilde E}
\def\tF{\widetilde F}
\def\tA{\widetilde A}
\def\im{{{\rm i}}}
\def\tY{{{\wtd Y}}}
\def\ep{{\epsilon}}
\def\vep{{\varepsilon}}
\def\bD{{{\bar D}}}
\def\R{{{\mathbb R}}}
\def\C{{{\mathbb C}}}
\def\H{{{\mathbb H}}}
\def\CP{{{\mathbb C}{\mathbb P}}}
\def\RP{{{\mathbb R}{\mathbb P}}}
\def\Z{{{\mathbb Z}}}
\def\bA{{{\mathbb A}}}
\def\bB{{{\mathbb B}}}
\def\bC{{{\mathbb C}}}
\def\bD{{{\mathbb D}}}
\def\bE{{{\mathbb E}}}
\def\bZ{{{\mathbb Z}}}
\def\Re{{{\frak{Re}}}}
\def\Im{{{\frak{Im}}}}
\def\cosec{{\,\hbox{cosec}\,}}
\def\Gm{{\Gamma_{\!\! -}}}
\def\Gp{{\Gamma_{\!\! +}}}
\def\stan{{standard }}
\def\nonstan{{supernumerary }}
\def\p{{\partial}}
\def\kdel#1{{\fft{\del}{\del#1}}}

\def\bog{{Bogomolny }}
\def\om{{\omega}}

\newcommand{\tamphys}{\it George and Cynthia Woods Mitchell  Institute
for Fundamental Physics and Astronomy,\\
Texas A\&M University, College Station, TX 77843, USA}

\newcommand{\auth}{
H. L\"u, Jianwei Mei and C.N. Pope
}

\begin{document}

\begin{flushright}
\hfill{
MIFP-08-06}\\
\end{flushright}

\begin{center}

{\large {\bf New Black Holes in Five Dimensions 
             
}}

\vspace{25pt}

\auth

\vspace{10pt}
{\tamphys}

\vspace{25pt}

\underline{ABSTRACT}

\end{center}

   We construct new stationary Ricci-flat metrics of cohomogeneity 2 in five 
dimensions, which generalise the Myers-Perry rotating 
black hole metrics by adding
a further non-trivial parameter.  We obtain them {\it via} a construction that
is analogous to the construction by Plebanski and Demianski 
in four dimensions of the most general type D metrics.  Limiting cases of
the new metrics contain not only the general Myers-Perry black hole with
independent angular momenta, but also the single rotation black ring
of Emparan and Reall.  In another limit, we obtain new static metrics
that describe black holes whose horizons are distorted lens spaces
$L(n;m)= S^3/\Gamma(n;m)$, where $m\ge n+2\ge 3$.  They are 
asymptotic to Minkowski spacetime factored by $\Gamma(m;n)$.  In the general
stationary case, by contrast,
the new metrics describe spacetimes with an horizon 
and with a periodicity condition on the time coordinate; these examples
can be thought of as five-dimensional analogues of the four-dimensional
Taub-NUT metrics.

\vspace{15pt}

\thispagestyle{empty}

\pagebreak
\setcounter{page}{1}

\tableofcontents

\addtocontents{toc}{\protect\setcounter{tocdepth}{2}}


\section{Introduction}

   A very considerable literature exists on the subject of exact solutions
in four-dimensional general relativity (see, for example, \cite{exact}).  
With the advent of higher-dimensional supergravity and string theory, it
has become important to extend the search for exact solutions to the arena
of higher dimensions.  This is important because these solutions
may have intrinsic significance within string theory or M-theory. 
An investigation of higher-dimensional solutions is also interesting
because low dimensions, such as four, may have special features that
no longer persist in higher dimensions.  

    An example of particular importance
concerns the uniqueness of black-hole solutions.  In four dimensions, there
exist very powerful theorems which establish, among other things, that $S^2$
is the only allowed topology for the event horizon of a black hole.  
Furthermore, uniqueness theorems have been established which demonstrate
that a four-dimensional black hole solution of the vacuum Einstein equations
is completely characterised by its mass and angular momentum.  It has been
known for some time that the notion of black-hole uniqueness is very
much weaker in higher dimensions. The most obvious illustration is provided
by the five-dimensional black ring solution \cite{empereal},
which has an event horizon whose topology is $S^2\times S^1$.  This 
contrasts with the $S^3$ horizon topology of the static spherically-symmetric
Schwarzschild-Tangherlini \cite{schw,tang} five-dimensional black hole, 
and its rotating
generalisation, which was found in \cite{myper}.   Many other ring-like
black holes have also been found in five dimensions (see, for example,
\cite{empereal2}, for a recent summary).

   There are also other simple examples of black holes with non-standard
geometry in $D\ge 5$ dimensions.  Starting from the usual cohomogeneity-1
Schwarzschild-Tangherlini metric, which is spherically symmetric and has
an $S^{D-2}$ horizon geometry, one can replace the round $(D-2)$-sphere
by any other Einstein manifold with the same Ricci curvature.  Of course
the metric will be asymptotically flat (in the strict sense of approaching 
Minkowski spacetime at infinity) only for the case of the round
$(D-2)$-sphere, but the curvature will go to zero at infinity for all
choices.  In five dimensions, the only options involve replacing
$S^3$ by $S^3/\Gamma$, where $\Gamma$ is some subgroup of $SO(4)$
that acts freely on $S^3$.  Examples are the lens spaces $L(m;n)=
S^3/\Gamma(m;n)$.  The spacetime is asymptotic to (Minkowski)$_5/\Gamma(m;n)$.
One of the results in this paper is to show that aside from these trivial
generalisations of the cohomogeneity-1 Schwarzschild-Tangherlini solution, 
there exist more complicated cohomogeneity-2 black holes that also
have (Minkowski)$_5/\Gamma(m;n)$ asymptotic structures.

    A particularly rich class of solutions in four dimensions is
provided by the Type D metrics.  These include the Schwarzschild and
Kerr black holes, and also the Taub-NUT metrics.  A convenient formulation
of many of the Type D metrics, including Kerr-Taub-NUT,
 was given by Plebanski in \cite{pleb}.  Subsequently, Plebanski and 
Demianski \cite{plebdemi} gave an elegant formulation of the most
general Type D metrics.  Amongst vacuum solutions, the key new feature
in the extension by Plebanski and Demianski was the inclusion of the
acceleration parameter as well as the mass, NUT charge and angular
momentum.  A cosmological constant can also be included.
(See \cite{exact} for a more complete discussion of these metrics.)

   The generalisation of the Kerr solution to arbitrary dimension was
obtained by Myers and Perry \cite{myper}.  This was later generalised to
include a cosmological constant; in five dimensions in \cite{hawhuntay}, and
in arbitrary dimensions in \cite{gilupapo1,gilupapo2}.  In further 
generalisations, it was shown in \cite{chenlupope1,chenlupope2} that
NUT charges can also be introduced in the higher-dimensional rotating
black hole solutions.  (The counting of parameters is different in even
and in odd dimensions.  In $D=2n$ dimensions there are $(n-1)$ 
independent NUT charges
and $(n-1)$ independent angular momenta,
while in $D=2n+1$ dimensions there are $(n-2)$ NUT charges and
$n$ angular momenta.)  In the process of 
constructing the NUT-charged solutions
\cite{chenlupope1,chenlupope2}, the metrics were
cast into a form that is a very natural higher-dimensional 
generalisation of the four-dimensional Plebanski metric of \cite{pleb}.

  It is natural now to investigate whether yet more general solutions 
can be obtained in higher dimensions by procedures that generalise
the construction given in \cite{plebdemi}.  In the case of five dimensions, 
we find that this can indeed be done, leading to a new larger class
of vacuum solutions.  Our new five-dimensional metrics contain one additional
non-trivial parameter, over and above the parameters in the Myers-Perry
black hole (making three non-trivial dimensionless parameters plus 
one scale parameter, in total).  

   After obtaining the new solutions, we give a detailed analysis of their
local and global properties.  First, we show by taking an appropriate
limit that they contain the Myers-Perry black holes (with two independent
rotation parameters) as a special case, and by taking a different limit,
we show that they contain the original black ring solution of
\cite {empereal} as another special case.  A third limit gives rise to
a class of static metrics.

   In a detailed analysis of the static metric limit, we show that 
with appropriate choices for the parameters these describe black-hole 
spacetimes in which the asymptotic spacetime geometry is (Minkowski)$_5/
\Gamma(m;n)$, where $\Gamma(m;n)$ is a certain discrete subgroup of
$SO(4)$.   The spatial surfaces at large radius are the lens space
$L(m;n)=S^3/\Gamma(m;n)$.  The horizon, on the other hand, has the
topology of 
the lens space $L(n;m)$ (with a distorted, non-Einstein metric).  
Note that these new solutions are very different from the five-dimensional
generalised Schwarzschild-Tangherlini ``lens space black holes'' mentioned 
earlier.  In those examples, which have cohomogeneity 1, both
the horizon and the spatial sections at infinity are round $L(m;n)$
spaces, whereas in our new metrics, which have cohomogeneity 2, there
is a sort of ``slumping'' in which the horizon has $L(n;m)$ topology
while the spatial sections at infinity have $L(m;n)$ topology.

   The difference between these lens space black holes can be seen by
looking at the dimensionless quantity $16\pi S T^4$.  This is equal
to $1/m$ for the factored Schwarzschild-Tangherlini black hole with
$L(m;n)$ asymptotic structure, and is larger than $1/m$ for our new
``slumped'' black hole with the same $L(m;n)$ asymptotic structure.

   We also study the global structure of the general, stationary,
metrics.  These
turn out to have properties that are somewhat analogous to those of
the four-dimensional Taub-NUT metrics, in that the time coordinate 
must be periodic in order to avoid conical singularities.  Unlike the
four-dimensional Taub-NUT metrics, however, there is no fibering of
the time coordinate at infinity.

\section{The General Local Solution}

\subsection{The new five-dimensional Ricci-flat metrics}\label{plebdem5sec}

   The five-dimensional vacuum solution describing a rotating black hole
with two independent angular momenta is contained within the
higher-dimensional rotating black holes found in \cite{myper}.  The 
generalisation of the five-dimensional solution to include a 
cosmological constant was obtained in \cite{hawhuntay}.  It was then shown 
in \cite{chenlupope1,chenlupope2} that this solution to the Einstein equations 
$R_{\mu\nu}= 4\lambda g_{\mu\nu}$ could be written in the simple
form 
\bea
ds_5^2 &=& \fft{x-y}{4X}\, dx^2 + \fft{y-x}{4Y}\, dy^2 + 
     \fft{X\, (d\phi + y d\psi)^2}{x(x-y)} + 
              \fft{Y\, (d\phi + x d\psi)^2}{y(y-x)} \nn\\
&& + \fft{a_0}{x y}\, \Big(d\phi + (x+y) d\psi + x y dt\Big)^2\,,
\label{pleb5}
\eea
where
\be
X= a_0 + a_1\, x + a_2 \, x^2 -\lambda x^3\,,\qquad
Y= a_0 + b_1\, y + a_2\, y^2 -\lambda y^3\,.\label{pleb5XY}
\ee
The constants $a_0$, $a_1$, $a_2$ and $b_1$ are related to the two angular
momenta, the mass and the NUT parameter \cite{chenlupope1,chenlupope2}.
Note that the metric has a coordinate scaling symmetry $x\rightarrow
\mu x$, $y\rightarrow \mu y$, 
which can be used to eliminate the NUT parameter \cite{chenlupope1}.
To interpret this metric as a rotating black hole, appropriate Wick
rotations must be performed.  In particular, $\phi$ is Wick rotated to
become the time coordinate.

   The metric (\ref{pleb5}) is closely analogous to the form of the
four-dimensional Kerr-de Sitter metric given in \cite{pleb}.  It is therefore 
natural to seek a generalisation of (\ref{pleb5}), analogous to the
four-dimensional generalisation with acceleration parameter
that was given in \cite{plebdemi}.  
Accordingly, we may try an ansatz of the form
\bea
ds_5^2 &=& \Omega_1(xy) \Big[\fft{x-y}{4X} dx^2 + \fft{y-x}{4Y} dy^2 +
     \fft{X\, (d\phi + y d\psi)^2}{x(x-y)} +
              \fft{Y\, (d\phi + x d\psi)^2}{y(y-x)}\Big] \nn\\
&& + \fft{a_0}{x y}\, \Omega_2(xy) 
   \Big(d\phi + (x+y) d\psi + x y dt\Big)^2\,,
\label{plebdem55}
\eea
where again we assume $X=X(x)$ and $Y=Y(y)$. 

   We find that provided we assume the cosmological constant vanishes,
Ricci-flat solutions of the form (\ref{plebdem55}) can arise, in the
case that
\be
\Omega_1(xy) = \fft1{(1-xy)^2}\,,\qquad \Omega_2(xy) = 1\,.
\ee
Specifically, we find that the metric
\bea
ds_5^2 &=& \fft1{(1-xy)^2}\Big[\fft{x-y}{4X} dx^2 + \fft{y-x}{4Y} dy^2 +
     \fft{X\, (d\phi + y d\psi)^2}{x(x-y)} +
              \fft{Y\, (d\phi + x d\psi)^2}{y(y-x)} \Big] \nn\\
&& + \fft{a_0}{x y}\, 
   \Big(d\phi + (x+y) d\psi + x y dt\Big)^2\,,
\label{plebdem5}
\eea
is Ricci flat, provided that $X$ and $Y$ are given by
\bea
X &=& a_0 + a_3\, x + a_2\, x^2 + a_1 \, x^3 + a_0\,  x^4\nn\nn\\
Y &=& a_0 + a_1\, y + a_2\, y^2 + a_3\, y^3 + a_0\, y^4\,.
\eea

   In what follows, it will sometimes turn out to be convenient to make a change
of coordinates in which we send 
\be
x\rightarrow 1/x\,,\quad t\rightarrow \im\, t\,,\quad
\phi\rightarrow \im\, \phi\,,\quad
 \psi\rightarrow \im\, \psi\,.
\ee
After doing so,
the metric (\ref{plebdem5}) can be written as
\bea
ds_5^2\!\!\! &=&\!\!\! \fft1{(x-y)^2}\Big[ \fft{x(1-xy)dx^2}{4 G(x)} -
     \fft{x(1-xy) dy^2}{4 G(y)} - \fft{G(x) (d\phi+y d\psi)^2}{(1-xy)}
  + \fft{x G(y)(d\psi + x d\phi)^2}{y(1-xy)}\Big]\nn\\
&& -\fft{a_0\, y}{x}\,\Big( dt + \fft{x}{y}\, d\phi + 
                            (x+ y^{-1}) d\psi\Big)^2\,,\label{plebdem51}
\eea
where
\be
G(\xi) \equiv a_0 + a_1\, \xi + a_2\, \xi^2 + a_3\, \xi^3 + a_0\, \xi^4\,.
\label{Gdef51}
\ee

  Since a Ricci-flat metric remains Ricci-flat when scaled by any constant
factor, one can absorb one of the four parameters in (\ref{Gdef51}) into
an overall dimensionful scale. This then implies that the local metric 
(\ref{plebdem51}) has three non-trivial continuous parameters.\footnote{In the
same sense, the Schwarzschild solution has no non-trivial continuous 
parameters, since the scale of the mass can be absorbed through an
overall rescaling of the metric.}  These three parameters are (by definition)
dimensionless.

   As we shall now show, the metric (\ref{plebdem5}) admits several
limiting forms that are of interest.

\subsection{Limits of the new five-dimensional metrics}
\label{limitssec}

   We find that there are three limiting cases that are of particular
interest:

\medskip
\noindent
\underline{{\bf Case I}}:
\medskip

   In this limit, we start from the general metric in the form (\ref{plebdem5})
and then send
\bea
&&x \rightarrow \ep^2\, x\,,\quad y\rightarrow \ep^2 \, y\,,\quad
 \phi\rightarrow \ep^{-1}\, \phi \,,
         \quad \psi\rightarrow \ep^{-3}\, \psi
\,,\quad t\rightarrow \ep^{-5}\,  t\,,\nn\\
&& a_0\rightarrow \ep^6 \, a_0\,,\quad a_1\rightarrow \ep^4 \, a_1\,,\quad
   a_2\rightarrow \ep^2 \, a_2\,,\quad
   a_3\rightarrow \ep^4 \, a_3\,.\label{caseI}
\eea
Upon sending $\ep$ to zero, the metric reduces to (\ref{pleb5}), with
\be
X= a_0 + a_1\, x + a_2\, x^2\,,\qquad Y= a_0 + a_3\, y + a_2\, y^2\,,
\label{XYsol}
\ee
which is of the form (\ref{pleb5XY}) with vanishing cosmological constant. 
Thus in the $\ep\rightarrow 0$ limit of (\ref{caseI}), the general metric
reduces to the five-dimensional Myers-Perry black hole, in the
form given in \cite{chenlupope1,chenlupope2}.  It has two non-trivial
continuous (dimensionless) parameters.  (One of the four parameters in
(\ref{XYsol}) can be absorbed by means of a coordinate transformation, and
a second by making an overall constant scaling of the metric.)

\medskip
\noindent
\underline{{\bf Case II}}:
\medskip

   To describe this limit, it is convenient to start from the general 
metric in the form (\ref{plebdem51}).  We then scale the coordinates and 
parameters according to
\bea
&&x\rightarrow \ep^2\, x\,,\quad y\rightarrow \ep^2\, y\,,\quad 
 \phi\rightarrow \ep\, \phi\,,\quad
\psi\rightarrow \ep\, \psi\,,\quad
  t\rightarrow \ep^{-1}\, t\,,\nn\\
&&a_0\rightarrow \ep^2\, a_0\,,\quad a_1\rightarrow a_1\,,\quad
a_2\rightarrow \ep^{-2}\, a_2\,,\quad a_3\rightarrow \ep^{-4}\, a_3\,.
\eea
   Upon sending $\ep$ to zero, this leads to the metric
\be
ds_5^2 = \fft1{(x-y)^2}\Big[ \fft{x dx^2}{4 G(x)} -
     \fft{x dy^2}{4 G(y)} - G(x) d\phi^2
  + \fft{x G(y)d\psi^2}{y}\Big]
 -\fft{a_0\, y}{x}\,\Big( dt +  y^{-1}\, d\psi\Big)^2\,,\label{ring1}
\ee
where
\be
G(\xi) \equiv a_0 + a_1\, \xi + a_2\, \xi^2 + a_3\, \xi^3\,.\label{ringG}
\ee
The local metric in this limit has two non-trivial
continuous (dimensionless) parameters.  (One of the four parameters in
(\ref{ringG}) can be absorbed by means of a coordinate transformation, and
a second by making an overall constant scaling of the metric.)

  As we shall discuss in appendix \ref{ringsec}, this metric contains the
original black ring, found in \cite{empereal}.  The {\it Case II} limit
is a five-dimensional analogue of the limit in which the Plebanski-Demianski
metric gives rise to the C-metric in four dimensions (see, for example,
\cite{exact}).  In fact, the local black ring solution was obtained from
Wick rotation of the Kaluza-Klein lifting \cite{9607236} of a dilatonic 
generalisation
of the four-dimensional C-metric \cite{9309075}.

\medskip
\noindent
\underline{{\bf Case III}}:
\medskip

   The third limiting form that we shall consider is obtained from
(\ref{plebdem51}) by making the scaling
\be
a_0\rightarrow\ep^2\, a_0\,,\qquad a_0^{1/2}\,t\rightarrow \ep^{-1}\, t\,,
\qquad \phi\rightarrow \im\phi\,,\qquad \psi\rightarrow\im \psi\,,
\ee
with all other coordinates and parameters left unscaled.  Upon sending
$\ep$ to zero, we obtain the metric
\bea
ds_5^2\!\!\!\! &=&\!\!\!\! \fft1{(x-y)^2}\Big[ \fft{(1-xy)dx^2}{4 G(x)} -
     \fft{x(1-xy) dy^2}{4y G(y)} + \fft{x G(x) (d\phi+y d\psi)^2}{(1-xy)}
  - \fft{x G(y)(d\psi + x d\phi)^2}{(1-xy)}\Big]\nn\\
&& -\fft{y}{x}\,dt^2\,,\label{bh5}
\eea
where
\be
G(\xi) \equiv  a_1 + a_2\, \xi + a_2\, \xi^2 \,.\label{Gbh}
\ee
Note that the Wick rotations of $\phi$ and $\psi$ in (\ref{bh5}) are 
performed just for later convenience; the same effect could be achieved
by sending $x\rightarrow -x$, $y\rightarrow - y$ and $\psi\rightarrow
-\psi$.  

   The metric (\ref{bh5}) has two non-trivial continuous (dimensionless)
parameters.  (One of the three parameters in (\ref{Gbh}) can be absorbed
by an overall constant scaling of the metric.)
As we shall discuss below, these static metrics describe
a rather wide class of black holes.

\section{Static Black Holes with New Geometry}\label{staticsec}

The local metric of the static black holes that we
are going to discuss is given by (\ref{bh5}). We choose to parameterise
the constants $a_i$ in such a way that the function $G$ becomes
\be
G(\xi) = -\mu^2 (\xi-\xi_1)(\xi-\xi_2)\,,
\ee
where 
\be
0<\xi_1\le \xi_2\,,\qquad
\xi_1 \xi_2 \le 1\,.
\ee
The coordinates $x$ and $y$ lie in the ranges 
\be
\xi_1\le x\le \xi_2\,,\qquad -\infty\le y\le \xi_1\,.
\ee
 The asymptotic
region at infinity occurs at $x=\xi_1=y$, and the horizon is located at
$y=0$.  There is a power-law singularity at $y=\infty$, which is
hidden by the horizon when the $\xi_i$ parameters
are chosen as described above.  (There would also be a power-law singularities
at $x=0$ and $xy=1$, but these do not lie within the spacetime manifold, for
the choice of coordinate ranges and parameters we are making.)
The metric contains no closed time-like circles outside the
horizon.

   Two special cases arise.  One case is
when $\xi_1 \xi_2=1$, for which the solution
reduces to standard five-dimensional Schwarzshild-Tangherlini black hole.  
The other special case 
is when $\xi_1=\xi_2$.  This gives rise to the Kaluza-Klein monopole,
and it is discussed in appendix \ref{KKmonosec}.   Our focus in this
section, therefore, will be when the parameters lie in the range
\be
0<\xi_1<\xi_2\,,\qquad \xi_1 \xi_2 <1\,.\label{xirange}
\ee

    To determine the periods that  the azimuthal coordinates $\phi$ and
$\psi$ must take in order to avoid any possible conical singularities,
we need to investigate the spacelike Killing vectors that degenerate
to zero length at each of the three locations
$x=\xi_1$, $x=\xi_2$ and $y=\xi_1$.  We normalise them by requiring
that each have unit Euclidean surface gravity at its corresponding 
degeneration surface.   (This ensures that each is associated with a 
$2\pi$ period.)  We find that the three degenerate 
Killing vectors are given by
\bea
x=\xi_1:&& \ell_1 =
\fft{\del}{\del \phi_1}\,,\nn\\
x=\xi_2:&& \ell_2 =\alpha\, \fft{\del}{\del\phi_1} +
\beta\, \fft{\del}{\del\phi_2}\,,\nn\\
y=\xi_1:&& \ell_3=\fft{\del}{\del\phi_2}\,,\label{statickilling}
\eea
where we have defined two new azimuthal coordinates $\phi_1$ and $\phi_2$,
related to the original $\psi$ and $\phi$ coordinates by
\be
\phi_1 = \fft{\mu^2 \sqrt{\xi_1} (\xi_2-\xi_1)
(\phi+ \xi_1 \psi)}{1-\xi_1^2}\,,\qquad
\phi_2 = \fft{\mu^2 \sqrt{\xi_1} (\xi_2-\xi_1)
(\psi +\xi_1 \phi)}{1-\xi_1^2}\,.
\ee
The constants $\alpha$ and $\beta$ in (\ref{statickilling}) are given by
\be
\alpha=\fft{(1-\xi_1\xi_2)\sqrt{\xi_1}}{(1-\xi_1^2)\sqrt{\xi_2}}\,,
\qquad
\beta=-\fft{(\xi_2-\xi_1)\sqrt{\xi_1}}{(1-\xi_1^2)\sqrt{\xi_2}}\,.
\label{alphabeta}
\ee

   It is clear that the Killing vectors $\ell_1,\ell_2$ and $\ell_3$ are
linearly dependent. In order to avoid conical singularities, it is
necessary that the coefficients of the linear dependence be
coprime integers, {\it i.e.}
\be
p \ell_1 + m \ell_2 + n\, \ell_3 =0\,.\label{ellcon0}
\ee
(See \cite{cvlupapo} for a discussion of this
technique for studying the removal of conical singularities in
metrics with degeneration surfaces.)
Furthermore, note that $\ell_2$ and $\ell_3$ can be simultaneously
degenerate when $x=\xi_2$ and $y=\xi_1$, which implies that
any linear combination of $\ell_2$ and $\ell_3$ is also a degenerate
Killing vector at this surface.
For the coprime integer pair $(m,n)$, the minimum period generated by
$m\ell_2 + n\ell_3$ is $2\pi$.  It follows that in order
to avoid a conical singularity, we must have $p=\pm 1$.  Without
loss of generality, let $p=-1$, and hence
\be
\ell_1=m \ell_2 + n \ell_3\,.\label{ellcon}
\ee
It follows from (\ref{statickilling}) and (\ref{alphabeta}) that we have
\be
\fft{(1-\xi_1^2)\sqrt{\xi_2}}{(1-\xi_1\xi_2)\sqrt{\xi_1}}=m\,,
\qquad
\fft{\xi_2-\xi_1}{1-\xi_1\xi_2}=n\,.\label{mnrel}
\ee
Thus the solution space is parameterised by a pair of coprime integers
$(m,n)$.  For the parameter range specified in (\ref{xirange}),
the integers $(m,n)$ must obey the inequalities
\be
m\ge n+2 \ge 3\,.\label{mnrange}
\ee
(The case $n=1$ occurs when $\xi_2=1$.)

  To understand the global structure of the spacetime, we first
examine the region at the double degeneration $(x=\xi_2,y=\xi_1)$ in more
detail.  It is useful to introduce two coordinates $\rho$ and $\vartheta$,
related to $x$ and $y$ by
\be
x=\xi_2 -\fft{\rho^2\, \sin^2\vartheta}{\xi_1}
     \,,\qquad y=\xi_1 - \fft{\rho^2\, \cos^2\vartheta}{\xi_2}\,.
\ee
The double degeneration will occur at $\rho=0$.  We also introduce
new azimuthal coordinates $\chi_1$ and $\chi_2$, defined by
\be
\chi_1= m\phi_1\,,\qquad \chi_2= \phi_2 + n\phi_1\,.\label{chidef}
\ee
These are defined so that the Killing vectors $\ell_2$ and $\ell_3$
defined in (\ref{statickilling}) are simply given by
\be
\ell_2=\fft{\del}{\del\chi_1}\,,\qquad 
\ell_3= \fft{\del}{\del\chi_2}\,.\label{l2l3}
\ee

   For small $\rho$, we then find that the metric (\ref{bh5})
approaches
\be
ds^2 = -\fft{\xi_1}{\xi_2}\, dt^2 +
  \fft{(1-\xi_1\xi_2)}{\mu^2\xi_1(\xi_3-\xi_1)^3}\,
\Big[ d\rho^2 + \rho^2(d\vartheta^2 + \sin^2\vartheta\, d\chi_1^2 +
        \cos^2\vartheta\, d\chi_2^2)\Big]\,.
\ee
In order not to have conical singularities, we see that $\chi_1$ and 
$\chi_2$ must independently have period $2\pi$. (In other words,
they are defined on a square lattice of side $2\pi$.)  This ensures that
the constant-$\rho$ surfaces at small $\rho$ are precisely round 3-spheres,
with no identifications.  The periodicities $\Delta\chi_1=2\pi$,
$\Delta\chi_2=2\pi$ are consistent with the expectation from (\ref{l2l3}).

   One might think that there would be another simultaneous degeneration
surface at $x=\xi_1=y$, which could lead to another periodicity
restriction.  This, however, is not the case.  As we already mentioned,
$x=\xi_1=y$ is actually asymptotic infinity.  This can be seen clearly if
we introduce new coordinates $r$ and $\theta$, defined in terms of $x$
and $y$ by
\be
\fft{\sqrt{\xi_1 - y}}{x-y} = \fft{\mu\sqrt{\xi_2-\xi_1}}{\sqrt{1-\xi_1^2}}
\,r\,\cos\theta\,,\qquad
\fft{\sqrt{x-\xi_1}}{x-y}=\fft{\mu\sqrt{\xi_2-\xi_1}}{\sqrt{1-\xi_1^2}}
\,r\,\sin\theta\,.\label{asympco}
\ee
As $r\rightarrow \infty$, the metric can be seen to approach Minkowski
spacetime locally, with
\be
ds^2=-dt^2 + dr^2 + r^2 (d\theta^2 + \sin^2\theta\, d\phi_1^2 +
\cos^2\theta\, d\phi_2^2)\,.\label{asymet}
\ee
Inverting (\ref{chidef}), we see that the azimuthal coordinates $\phi_1$
and $\phi_2$ are related to $\chi_1$ and $\chi_2$ by
\be
\phi_1= \fft1{m}\, \chi_1\,,\qquad \phi_2 = \chi_2 - \fft{n}{m}\, \chi_1\,.
\label{phifromchi}
\ee

   Our discussion of the regularity conditions at the double degeneration
$(x=\xi_2,y=\xi_1)$ showed that $\chi_1$ and $\chi_2$ are periodic on a 
square lattice of side $2\pi$.  It then follows from (\ref{phifromchi})
that $\phi_1$ and $\phi_2$ are periodic on a tilted lattice, in which
identifications are made under the two operations
\bea
1)\qquad && \phi_1\longrightarrow \phi_1\,,\qquad 
        \phi_2\longrightarrow \phi_2 + 2\pi\,,\nn\\
2)\qquad && \phi_1\longrightarrow \phi_1 + \fft{2\pi}{m}\,,\qquad
 \phi_2\longrightarrow \phi_2 - \fft{2\pi n}{m}\,.\label{Lmn}
\eea
These are precisely the identifications that arise for the lens space
$L(m;n)$.  This can be seen from the definition of $L(m;n)$.  One takes
$S^3\subset \C^2$, with complex coordinates $(z_1,z_2)$, and quotients
according to 
\be
(z_1,z_2) \equiv (z_1\, e^{2\pi \im/m},z_2\, e^{2\pi\im  n/m})\,,
\ee
(where $m$ and $n$ are coprime integers with $1\le n\le m-1$.)
Taking $z_1= \sin\theta \, e^{\im \phi_1}$ and $z_2=\cos \theta\,
 e^{-\im \phi_2}$, we see that the lens space $L(m;n)$ is indeed
defined by the identifications (\ref{Lmn}).  

       The horizon of the black hole is located at $y=0$, and 
from (\ref{bh5}), its metric is given by
\bea
ds_H^2 &=&
\fft{(x-\xi_1)\Big( \xi_2 (1-\xi_1^2) -  x (1-\xi_1 \xi_2)\Big)}
{\mu^2 \xi_1 (\xi_1 - \xi_2)^2 x} \Big(d\phi_1  +
\fft{\xi_1 (1-\xi_1\xi_2)}{\xi_2 \Big((1-\xi_1^2) - x(1-\xi_1 \xi_2)\Big)}
d\phi_2\Big)^2\nn\\
&&+\fft{(1-\xi_1^2)\xi_2(\xi_2-x)}{\mu^2 (\xi_2-\xi_1)^2
x\Big(\xi_2(1-\xi_1^2) - x (1-\xi_1 \xi_2)\Big)}d\phi_2^2+
\fft{dx^2}{4x^2 G(x)}\,.\label{hormet1}
\eea
It is easy to verify that this is not an Einstein metric, and
it is not homogeneous.

   In order to understand the geometry of the event horizon, it is
helpful to introduce new azimuthal coordinates $\td\phi_1$ and
$\td\phi_2$, chosen so that the Killing vectors $\ell_1$ and $\ell_2$
that vanish at $x=\xi_1$ and $x=\xi_2$ are simply given by
\be
\ell_1\equiv\fft{\del}{\del\td \phi_1}\,,\qquad
\ell_2\equiv\fft{\del}{\del\td \phi_2}\,.
\ee
These coordinates are related to $\chi_1$ and $\chi_2$, and to $\phi_1$
and $\phi_2$, by
\be
\td\phi_1= \fft1{n}\chi_2 = \phi_1 + \fft1{n} \phi_2\,,\qquad
\td\phi_2 = \chi_1-\fft{m}{n} \chi_2 = -\fft{m}{n}\phi_2\,.
\label{tdphidef}
\ee

   In terms of $\td\phi_1$ and $\td\phi_2$, the metric (\ref{hormet1}) 
on the horizon can conveniently be written in the following two ways:
\bea
ds_H^2\!\!\! &=&\!\!\! \fft{dx^2}{4\mu^2 x^2(\xi_2-x)(x-\xi_1)} +
  \fft{(\xi_2-x) g_1  \big(d\td\phi_2+ f_1 d\td\phi_1\big)^2 
}{\mu^2 x \xi_2(\xi_2-\xi_1)^2} + 
     \fft{(x-\xi_1) \xi_2}{\mu^2 x g_1}\, d\td\phi_1^2\,,\label{hor1}\\
ds_H^2 \!\!\!&=&\!\!\! \fft{dx^2}{4\mu^2 x^2(\xi_2-x)(x-\xi_1)} +
  \fft{(x-\xi_1) g_2  \big(d\td\phi_1+ f_2 d\td\phi_2\big)^2
}{\mu^2 x \xi_1(\xi_2-\xi_1)^2}  +
     \fft{(\xi_2 -x) \xi_1}{\mu^2 x g_2}\, d\td\phi_2^2\,,\label{hor2}
\eea
where
\bea
g_1 &=& (x-\xi_1) + (\xi_2-x)\xi_1\xi_2\,,\qquad 
  f_1 = \fft{(x-\xi_1)(1-\xi_1\xi_2)\sqrt{\xi_2} }{g_1\sqrt{\xi_1}}\,,\nn\\
g_2 &=& (\xi_2-x) + (x-\xi_1)\xi_1\xi_2\,,\qquad
  f_2 = \fft{(\xi_2 -x)(1-\xi_1\xi_2)\sqrt{\xi_1} }{g_2\sqrt{\xi_2}}\,.
\eea
We can study the regions in the vicinity of the ``north and south poles''
at $x=\xi_1$ and $x=\xi_2$ by defining a new ``latitude'' coordinate $\rho_1$
such that $x=\xi_1 +\rho_1^2$, or a new latitude coordinate $\rho_2$ such
that $x=\xi_2-\rho_2^2$.  Near $\rho_1=0$ at the north pole the metric
approaches
\be
ds_H^2 \rightarrow \fft1{\mu^2 \xi_1^2(\xi_2-\xi_1)}\,\Big( d\rho_1^2 + 
    \rho_1^2 d\td\phi_1^2\Big) + \fft1{\mu^2} d\td\phi_2^2\,,
\ee
whilst near $\rho_2=0$ at the south pole the metric approaches
\be
ds_H^2 \rightarrow \fft1{\mu^2 \xi_2^2(\xi_2-\xi_1)}\,\Big( d\rho_2^2 +
    \rho_2^2 d\td\phi_2^2\Big) + \fft1{\mu^2} d\td\phi_1^2\,.
\ee

   As far as degenerations of the local metric are concerned, $ds_H^2$ exhibits
the same essential behaviour as the standard metric on $S^3$,
\be
d\Omega^2_3 = d\theta^2 + \sin^2\theta d\td\phi_1^2 + 
         \cos^2\theta d\td\phi_2^2\,.\label{S3met}
\ee
The geometric {\it details} of the actual horizon metric $ds_H^2$ differ
from (\ref{S3met}) in several respects, but these are all in the form
of smooth distortions that do not have any impact on global topological
considerations. As we already remarked, the horizon metric is neither
Einstein nor homogeneous.

   It can be seen from the equations in (\ref{tdphidef}) that the 
relation between $(\td\phi_1,\td\phi_2)$ and $(\chi_1,\chi_2)$ is
just like the relation (\ref{phifromchi}) between $(\phi_1,\phi_2)$
and $(\chi_1,\chi_2)$, except that the roles of the integers $m$ and
$n$ are reversed.  It follows that if we now repeat, on the horizon, 
 the argument that
showed the topology of the $r=$constant spatial surfaces at large
$r$ are lens spaces $L(m;n)$, we will find that the topology of the
horizon is the lens space $L(n;m)$.  However, as mentioned above,
geometrically, the horizon is an inhomogeneously-distorted $L(n;m)$
lens space.  In view of the inequalities satisfied by $m$ and $n$,
which are given in (\ref{mnrange}), there are only $n$ inequivalent 
topologies for the horizons, since the lens spaces
$L(n;p)$ and $L(n;p+n)$ are identical.  Note that $n=1$ is an allowed 
value (see (\ref{mnrange})), in which case the lens spaces $L(1;m)$ are all
topologically just $S^3$.

      It is now a straightforward matter to calculate the area $A$ of the
horizon, and hence its entropy $S=\ft14 A$.  It is given by
\bea
S&=&\ft14 \int \sqrt{g_3} = \ft14 \int d\phi_1 d\phi_2
\int_{\xi_1}^{\xi_2} dx \fft{(1-\xi_1^2)\sqrt{\xi_2}}{
2\mu^3 \sqrt{\xi_1}(\xi_2-\xi_1)^2 x^2}\nn\\
&=&\fft{\pi^2 (1-\xi_1\xi_2)}{2\mu^3 \xi_1 \xi_2 (\xi_2- \xi_1)}\,.
\label{hent}
\eea
We may also calculate the surface gravity $\kappa$, 
computed for the timelike Killing vector $K_0=\del/\del t$, and hence obtain
the Hawking temperature $T=\kappa/(2\pi)$.  It is given by
\be
T=\fft{\mu \sqrt{\xi_1\xi_2}}{2\pi}\,.\label{htemp}
\ee

 The ADM mass is also easy to calculate, by means of a Komar integral.
It is given by
\be
M=\fft{3}{32\pi} \int *dK_0 =\fft{3\pi (1-\xi_1 \xi_2)}{8\mu^2
(\xi_2-\xi_1)\sqrt{\xi_1\xi_2}}\,.\label{bhmass}
\ee
It should be noted that in this calculation, involving an integration
over the boundary $L(m;n)$ lens space at infinity, and also in the
calculation of the
 horizon area, involving an integration over the
inhomogeneous lens space $L(n;m)$ at the horizon, one must take care
to handle the azimuthal coordinate integrations carefully, paying due
regard to the periodicity conditions implied by the lens-space 
identifications.  The general rule is that when a given 3-sphere metric
is factored to give the lens space $L(p;q)$, the 3-volume is reduced by
a factor of $1/p$.

   It is straightforward to verify that the black holes satisfy the
first law of thermodynamics, namely
\be
dM=T\, dS\,.
\ee
Furthermore, we have
\be
M=\ft32 T\,S\,,
\ee
as in the case of the standard Schwarzschild black hole in five
dimensions.

       Thus we have constructed a large class of $D=5$ static
black holes whose topology is specified by a pair of
integers $(m,n)$ lying in the range of (\ref{mnrange}).
The metric has three
linearly-dependent degenerate spacelike Killing vectors 
$(\ell_1,\ell_2,\ell_3)$
with unit Euclidean surface gravity.  These Killing vectors slump
from asymptotic infinity to the horizon.  In the horizon, there are
only two degenerate Killing vectors $(\ell_1,\ell_2)$, giving
rise to a geometry of non-homogeneously distorted
lens space $L(n;m)$.   In the asymptotic region,
the (only) two degenerate Killing vectors are $(\ell_1,\ell_3)$,
and the large-$r$ spatial sections have the geometry of homogeneous
lens spaces $L(m;n)$.

   It should be emphasised that these results do not contradict results
on the uniqueness of higher-dimensional static asymptotically-flat black
holes in \cite{gibidashi}.  Since the spatial sections at large distance
in our new solutions have the topology of the $L(m;n)$ lens space, which
is the quotient of $S^3$ by a certain discrete subgroup $\Gamma(m;n)$
of $SO(4)$,
it follows that although the curvature tends to zero at infinity the
spacetime is not asymptotic to Minkowski spacetime, but, rather, to
the quotient Minkowksi/$\Gamma(m;n)$.   Thus the conditions assumed
in \cite{gibidashi}, under which uniqueness could be proved, are not
satisfied.

   One can also, of course, consider a different and considerably
simpler static black hole with the same asymptotic geometry
Minkowksi/$\Gamma(m;n)$.  As was noted in \cite{gibidashi}, the 
round $S^n$ in any $D=n+2$ dimensional Schwarzschild-Tangherlini
solution can be replaced by an arbitrary Einstein space of the same
Ricci curvature.  Although the five-dimensional example was not discussed
explicitly in \cite{gibidashi}, one can simply replace $S^3$ in the
five-dimensional Schwarzschild-Tangherlini spacetime by the lens space
$L(m;n)$.  In this case, unlike our new solutions,
the horizon will have the same round $L(p;q)$ lens space geometry
as the large$-r$ spatial sections.
There are only two zero-length Killing vectors in the whole metric. 
These factored Schwarzschild-Tangherlini solutions are of cohomogeneity 1,
in contrast to our new solutions, which have cohomogeneity 2.
For each of the new solutions with asymptotic
$L(m;n)$ spatial sections that we have obtained in this paper, there is
another, inequivalent, black hole with the same asymptotic structure,
obtained instead by simply factoring the $S^3$ in the 
Schwarzschild-Tangherlini solution by $\Gamma(m;n)$.

    One way to compare the different black-hole metrics is to look at
the dimensionless quantity obtained by multiplying the entropy by the
cube of the temperature.  From (\ref{mnrel}), (\ref{hent}) and (\ref{htemp}) 
we find
\be
S= \fft{1}{16\pi T^3}\,  \fft{\sqrt{\xi_1\xi_2}}{n}\,.\label{STslump}
\ee
At fixed temperature, therefore, the entropy is {\it maximised} by the
Schwarzschild-Tangherlini spacetime, which corresponds to $m=n=1$ and
$\xi_1 \xi_2=1$.  It is interesting to note that the 
``factored Schwarzschild-Tangherlini''
solution, in which $S^3$ surfaces are quotiented to give
$L(m;n)=S^3/\Gamma(m;n)$, will have a smaller entropy than our new ``slumped''
black hole with $L(n;m)$ horizon topology.  This follows from the fact that
the former will have entropy $S=1/(16m\pi T^3)$, whereas the slumped solution
has entropy given by (\ref{STslump}), which is larger by the factor
\be
1 + \fft{\xi_1\, (1-\xi_1 \xi_2)}{\xi_2-\xi_1} \ge 1\,.
\ee

   One further remark concerns the limit $\xi_1\xi_2\rightarrow1$, 
which gives the usual Schwarzschild-Tangherlini metric.  It might appear
that the mass formula (\ref{bhmass}) is incompatible with this limit,
since it vanishes when $\xi_1\xi_2=1$.
To resolve this apparent paradox, we note that when $\xi_1 \xi_2=1$,
it follows from (\ref{statickilling}) that $\ell_2=-\del/\del\phi_2$,
Thus (\ref{ellcon0}) can be simply solved by letting
$p=0$ and $m=n=1$.  Then the condition (\ref{ellcon}) no longer
holds, and $\phi_1$ and $\phi_2$ both have independent
$2\pi$ periods.  The solution indeed describes the standard
Schwarzschild black hole.  However, within our general class of black-hole 
solutions, taking the limit $\xi_1\xi_2\rightarrow 1$ assumes that
the condition  (\ref{ellcon}) is still imposed.
This corresponds to sending $m$ and $n$ to infinity, while keeping
$m/n\rightarrow 1$.  The resulting metric then describes
a Schwarzshild-Tangherlini black hole in which the round $S^3$ is replaced by
$S^3/\Gamma(\infty;\infty)$.  This has zero volume, and so the mass would
vanish too.

\section{Charged Static Black Hole with New Geometry}

        Having obtained the new static black hole solutions, which 
exhibit the feature of having an $L(n;m)$ lens-space topology on the horizon,
which ``slumps'' to give $L(m;n)$ lens-space spatial sections at infinity,
we can easily construct charged generalisations, by using solution-generating 
techniques involving Kaluza-Klein reduction and U-duality.  
We shall consider charged solutions in $D=5$, $N=2$
supergravity coupled to two vector multiplets.  This $U(1)^3$ theory
can also be obtained as a truncation of the maximal $N=8$ supergravity
in $D=5$.  The bosonic Lagrangian is given by
\be
{\cal L}=\sqrt{-g}(R -\ft12 \sum_{i=1}^3 X_i^{-2} (\del X_i)^2-
\ft14 \sum_{i=1}^3 X_i^{-2} (F^i)^2 +
\ft14 \epsilon^{\mu\nu\rho\sigma\lambda} F_{\mu\nu}^1
F_{\rho\sigma}^2 A_\lambda^3\,,
\ee
where $F^i=dA^i$, and $X_1$, $X_2$ and $X_3$ satisfy $X_1X_2X_3=1$;
they describe two scalar fields.  

   By a standard solution-generating procedure involving lifting to six
dimensions, boosting, reducing and acting with U-duality, our previous
neutral static solution can be transformed into a charged one.
   This new charged black hole is given by
\bea
ds_5^2 &=& - (H_1 H_2 H_3)^{-\ft23}\, \fft{y}{x} dt^2 +
(H_1 H_2 H_3)^{\ft13}\Big\{\fft1{(x-y)^2}\Big[ \fft{(1-xy)dx^2}{4 G(x)} -
     \fft{x(1-xy) dy^2}{4y G(y)}\nn\\
&& \qquad\qquad + \fft{x G(x) (d\phi+y d\psi)^2}{(1-xy)}
  - \fft{x G(y)(d\psi + x d\phi)^2}{(1-xy)}\Big]\Big\}\,,\\
X_i&=&H_i^{-1}(H_1H_2H_3)^{1/3}\,,\quad
A=(1-H_i^{-1})\coth\beta_i\,dt\,,\quad
H_i=1 + \fft{\sinh^2\beta_i (x-y)}{x}\,,\nn
\eea
with $G(\xi)$ having the same form
\be
G(\xi)= -\mu^2 (\xi-\xi_1)(\xi-\xi_2)
\ee
as in the previous section.  

   The
global analysis proceeds in the same way as in the static
case.  We take the roots of $G(\xi)$ to satisfy the inequalities
$0<\xi_1<\xi_2$ and $\xi_1 \xi_2<1$, and the ranges of $x$ and $y$ is
the same as in the static case, namely $\xi_1\le x\le \xi_2$ and $-\infty
\le y\le \xi_1$.   Power-law
singularities are again avoided outside the horizon, since the functions 
$H_i$ are
positive definite outside the horizon.  

   The solution describes charged static black holes, which, as in the 
uncharged case, ``slump'' from a lens space topology $L(m;n)$ on the
spatial sections at infinity to $L(n;m)$ topology at the horizon.

   The mass, entropy, charge and their respective potentials can be
easily obtained, and are given by
\bea
M &=& \fft{\pi (3 + 2(s_1^2 + s_2^2 + s_3^2))(1-\xi_1\xi_2)}{
8\mu^2 \sqrt{\xi_1\xi_2}(\xi_2-\xi_1)}\,,\nn\\
S &=& \fft{\pi^2c_1c_2c_3(1-\xi_1\xi_2)}{
2\mu^3\xi_1\xi_2(\xi_2-\xi_1)}\,,\qquad
T=\fft{\mu\sqrt{\xi_1\xi_2}}{2\pi c_1c_2c_3}\,,\nn\\
Q_i&=&\fft{\pi c_i s_i (1-\xi_1\xi_2)}{4\mu^2\sqrt{\xi_1\xi_2}(\xi_2-\xi_1)}
\,,\qquad \Phi_i=\fft{s_i}{c_i}\,.\label{chargeth}
\eea
where $s_i=\sinh\beta_i$ and $c_i=\cosh\beta_i$.  These quantities
satisfy the expected thermodynamic relations
\be
dM=TdS + \Phi_i dQ_i\,,\qquad
M=\ft32 TS + \Phi_i Q_i\,.
\ee

   The solution can be straightforwardly lifted to six dimensions, where
it becomes a dyonic string with a pp-wave propagating along 
the sixth direction.

   The BPS limit, corresponding to $T=0$ and $\Phi_i=1$, with
\be
M=Q_1 + Q_2 + Q_3\,,
\ee
can be achieved by letting $\mu$ and $\beta_i$ approach infinity,
whilst keep the ratio $q_i=s_i/\mu$  fixed.   To implement
this limit in the solution, it is necessary to let $x$ approach $y$,
accompanied by an appropriate rescaling to keep the metric from degenerating
in this limit. 
In particular, we make the coordinate transformation (\ref{asympco}),
then set $s_i=q_i \mu$, and take the $\mu\rightarrow \infty$
limit.  The resulting metric is given by
\be
ds^2=- (H_1 H_2 H_3)^{-\ft23}dt^2 +
(H_1 H_2 H_3)^{\ft13} (dr^2 + r^2 d\Omega_3^2)\,,
\ee
where $H_i=1 + q_i^2(1-\xi_1^2)/(\xi_1(\xi_2-\xi_2) r^2)$, and
$d\Omega_3^2$ is the metric of the ``round'' lens space $L(m;n)$
(\ie it is an Einstein metric, obtained by factoring the unit 3-sphere
by the discrete subgroup $\Gamma(m;n)$ of $SO(4)$).
Thus, the ``slumping'' feature of the non-extremal solution
is lost in the extremal limit.  The mass and charge for
the extremal solution are simply those obtained from the expressions in
(\ref{chargeth}), upon taking the limit.  However, the entropy is changed,
since now the horizon has the same topology and geometry 
as the asymptotic spatial metric $d\Omega_3^2$.
It would be of interest to see whether one could take a different BPS limit 
of the non-extremal solution that retained the slumping feature.

\section{Global Analysis of the General Solutions}

  We now turn to an analysis of the full metric (\ref{plebdem51}) that we
found in section \ref{plebdem5sec}.  In 
section \ref{limitssec}, we saw that it admits a limit, described in 
{\it Case I}, in which it
becomes the standard Myers-Perry black hole with two independent 
rotation parameters.  Actually, the {\it Case I} limit ostensibly has
a further parameter in addition to the mass and the two angular
momenta, which one might wish to identify as a five-dimensional
NUT charge.  However, as discussed in \cite{chenlupope1,chenlupope2},
this ``NUT charge'' is really a trivial parameter, in the sense that
it can be removed by means of a coordinate transformation.

   In the full metric (\ref{plebdem51}), no analogous coordinate
transformation can be made, and so the additional parameter that becomes
the (trivial) NUT parameter in the {\it Case I} limit is now 
instead non-trivial.
We shall analyse the global structure of the full metric (\ref{plebdem51})
in this section.  Our findings are that for suitable choices of the
parameters we can obtain metrics that extend smoothly onto spacetime
manifolds that have horizons, but are otherwise free of conical singularities.
In order to achieve this, it is necessary for the time coordinate to
be appropriately periodically identified.  In this respect, the situation
is reminiscent of the Taub-NUT metric in four dimensions. However, there
are significant differences too, which will emerge as our discussion
proceeds.

   It is convenient to reparameterise the metric function $G$ in 
(\ref{Gdef51}) in the form
\be
G(\xi_)= \mu^2 (\xi - \xi_1) (\xi-\xi_2) (\xi-\xi_3) (\xi-\xi_4)\,,
\ee
where $a_0=\mu^2$ and  
\be
\xi_4 =\fft1{\xi_1 \xi_2 \xi_3}\,.  
\ee
The constant $\mu$ has dimensions
(length)$^{-1}$, whilst $\xi_1$, $\xi_2$ and $\xi_3$ are the three
non-trivial dimensionless parameters.  

    For reasons that will become
apparent later, shall restrict the parameters so that 
\be
\xi_1<-1<\xi_2< 0<\xi_3\le \xi_4\,, \qquad \hbox{and}\qquad 
         \xi_1\xi_2 \le 1\,.
\ee
The coordinates $x$ and $y$ will lie in the ranges 
\be
\xi_1\le x\le \xi_2\,,\qquad  \xi_2 \le y\le \infty\,.
\ee
The asymptotic region is approached at $x=y=\xi_2$, 
and the outer and inner horizons
are at $y=\xi_3$ and $y=\xi_4$ respectively.  The surface of the
ergosphere is at
$y=0$.   The curvature has power-law singularities at $xy=1$ and 
at $y=\infty$.  The former does not lie in the spacetime manifold,
and the latter lies behind the horizons.  For later convenience, 
we introduce
positive parameters 
\be
\eta_1\equiv -\xi_1\, \qquad \eta_2\equiv -\xi_2\,.
\ee

        We now analyse the conditions under which there are no
conical singularities outside the horizon.  To do this, it is useful
first to make coordinate transformations as follows.  We begin with
a redefinition of the time coordinate,
\be
t\rightarrow t/\mu + (\eta_2 + \eta_2^{-1}) \psi - \phi\,,
\ee
and then we introduce new azimuthal angles $\phi_1$ and
$\phi_2$, defined by
\bea
\phi_1 &=& \fft{\mu^2(\eta_1-\eta_2)(\eta_2 +\xi_3)
(1+\eta_1\eta_2^2\xi_3)}{\eta_1\eta_2^{3/2}\xi_3(1-\eta_2^2)}
(\phi-\eta_2\psi)\,,\nn\\
\phi_2 &=& \fft{\mu^2(\eta_1-\eta_2)(\eta_2 +\xi_3)
(1+\eta_1\eta_2^2\xi_3)}{\eta_1\eta_2^{3/2}\xi_3(1-\eta_2^2)}
(\psi-\eta_2\phi)\,,
\eea

    The metric degenerates at $x=-\eta_1$, $x= -\eta_2$ and 
$y=-\eta_2$.
The three corresponding degenerating Killing vectors, normalised
to have unit Euclidean
surface gravity, are given by
\bea
x=-\eta_2:&& \ell_1 = \fft{\del}{\del\phi_1}\,,\nn\\
y=-\eta_2:&&  \ell_2=\fft{\del}{\del\phi_2}\,,\\
x=-\eta_1:&& \ell_3=-\fft{\eta_1^{3/2}\xi_3(1-\eta_1\eta_2)}{\mu
(\eta_1+\xi_3)(1+\eta_1^2\eta_2\xi_3)}\fft{\del}{\del t}\nn\\
&&
+\fft{\sqrt{\eta_1}(\eta_2+\xi_3)(1+\eta_1\eta_2^2\xi_3)}{
\sqrt{\eta_2}(\eta_1+\xi_3)(1-\eta_2^2)(1+\eta_1^2\eta_2\xi_3)}\Big(
(1-\eta_1\eta_2)\fft{\del}{\del\phi_1}+
(\eta_1-\eta_2)\fft{\del}{\del\phi_2}\Big)\,.\nn
\eea
Since each $\ell_i$ independently generates a $2\pi$ translation
around its degeneration surface, it follows in particular 
that the time coordinate $t$ must be periodic; a property of any 
Taub-NUT solution.
Note that the $\del/\del t$ term in $\ell_3$ is absent if 
$\eta_1\eta_2=1$; this special case describes the Myers-Perry 
rotating black hole (and has no time periodicity).

    Now let us consider the asymptotic region, located at
$x=y=\xi_2=-\eta_2$.  We make the coordinate transformation
\be
\fft{\sqrt{\xi_2-x}}{y-x} = A\, r\cos\theta\,,\qquad
\fft{\sqrt{y-\xi_2}}{y-x} = A\, r \sin\theta\,,
\ee
where
\be
  A^2= \fft{\mu^2(\eta_1-\eta_2)(\eta_2+\xi_3)(1+\eta_1\eta_2^2 \xi_3)}{
   \eta_1\eta_2^2 \xi_3 (1-\eta_2^2)}\,.
\ee
Taking the limit $r\rightarrow \infty$, we see that the metric
at infinity approaches
\be
ds=-dt^2 + dr^2 + r^2 (d\theta^2 + \cos^2\theta d\phi_1^2 +
                      \sin^2\theta d\phi_2^2)\,.
\ee
From the form of the metric near infinity, the Komar integrals giving
the ADM mass and the angular momenta can be evaluated.  We find
\bea
M &=&\fft{3\pi \eta_1\eta_2\xi_3 (1-\eta_2^2)}{8\mu^2 
(\eta_1-\eta_2)(\eta_2+\xi_3)
(1+\eta_1\eta_2^2\xi_3)}\,,\nn\\
J_{\phi_2} &=& \fft{\pi\eta_1^2\eta_2^{3/2}\xi_3^2(1-\eta_2^2)^2}{
4\mu^3 (\eta_1-\eta_2)^2 (\eta_2+\xi_3)^2 (1+\eta_1\eta_2^2\xi_3)^2}
\,,\qquad
J_{\phi_1} = \eta_2\, J_{\phi_2}\,.
\eea

      The outer horizon is at $y=\xi_3$.   The asymptotically
timelike Killing vector that 
degenerates there is given by
\be
\ell_0=\fft{\del}{\del t} -
\fft{\mu(\eta_1-\eta_2)(1 + \eta_1\eta_2^2 \xi_3)}{
\eta_1\sqrt{\eta_2} (1+\eta_2\xi_3)(1-\eta_2^2)}
\Big((1+\eta_2\xi_3)\fft{\del}{\del\phi_2} -
(\eta_2+\xi_3)\fft{\del}{\del\phi_1}\Big)\,.
\ee
Calculating the surface gravity, and the volume of the horizon,
we obtain the temperature and the entropy,
given by
\bea
T&=&\fft{\mu(\eta_1 + \xi_3) (1-\eta_1\eta_2\xi_3^2)}{2\pi
\eta_1\sqrt{\xi_3} (1+ \eta_2\xi_3)}\,,\nn\\
S&=& \fft{\pi^2 (\eta_1\eta_2)^2\xi_3^{3/2} (1 + \eta_2\xi_3)(1-\eta_2^2)}
{2\mu^3 (\eta_1 + \xi_3)(\eta_1-\eta_2)(\eta_2+\xi_3)^2 (
1+\eta_1\eta_2^2\xi_3)^2}\,.
\eea
The angular velocities on the horizon are given by
\bea
\Omega_{\phi_1}&=&\fft{\mu(\eta_1-\eta_2)(\eta_2+\xi_3)(1+\eta_1
\eta_2^2\xi_3)}{\eta_1\sqrt{\eta_2} (1+\eta_2\xi_3)(1-\eta_2^2)}
\,,\nn\\
\Omega_{\phi_2}&=&-\fft{\mu(\eta_1-\eta_2)(1+\eta_1\eta_2^2\xi_3)}{
\eta_2\sqrt{\eta_2}(1-\eta_2^2)}\,.
\eea

   The first law of thermodynamics is not satisfied in these
solutions in general. In fact the analysis of the
thermodynamics of Taub-NUT solutions is notoriously unsettled.
However, a special case arises if $\eta_1\eta_2= 1$.
It is easy to verify in this case that we have
\be
dM=TdS + \Omega_{\phi_1} dJ_{\phi_1} + \Omega_{\phi_2} dJ_{\phi_2}
\,,\qquad
M=\ft32 (TS + \Omega_{\phi_1} J_{\phi_1} + \Omega_{\phi_2} J_{\phi_2}
)\,.
\ee
In fact when $\eta_1 \eta_2=1$, the metric is nothing but the Myers-Perry
rotating black hole, in an unusual coordinate system.

   It is interesting to note that the general local metric
(\ref{plebdem51}) gives rise to the Myers-Perry solution in
two very different limiting ways. One way is {\it via} the limit discussed
in {\it Case I} in section \ref{limitssec},  and the other is by taking
$\eta_1\eta_2=1$ in the general metrics.  The coordinate transformation
that links one to the other is quite complicated, and we have verified their
equivalence by studying the relationship between the
mass, entropy and angular momenta.

     Finally we remark that although the spacetimes we have obtained
here in five dimensions are in some respects similar to four-dimensional
Taub-NUT spacetimes, in that the time coordinate is periodic,
there are also significant differences.
  In four dimensions, it is a fibration in the time
direction at asymptotic infinity that is responsible
for imposing the periodicity of the time coordinate.  In our 
five-dimensional solution, on the other hand,
the metric approaches Minkowski spacetime locally at infinity,
with no fibration in the time direction.  Our solution is also very different
in structure from the topological soliton ``time machines'' 
obtained in \cite{hgwells},
where there are no horizons or singularities in the spacetime.
By contrast, our spacetimes here describe black objects 
with horizons and with singularities inside
the horizons.  Note, further, that the outer horizon of in solution,
which is located at $y=\xi_3$, is separated from the velocity of light surface
surrounding the time machine at $y=\xi_2=-\eta_2$.

\section{Conclusions}

     In this paper, we have constructed new stationary cohomogeneity 2 
solutions of the vacuum Einstein
equations in five dimensions.  We obtained it by starting from the 
five-dimensional
rotating black hole, written in the very simple form found in 
\cite{chenlupope1}, and then making an ansatz that involved generalising 
certain metric functions in the rotating black hole, and also introducing
a conformal factor in a four-dimensional subspace.  Our procedure is 
somewhat analogous to one that was performed in four dimensions in 
\cite{plebdemi}.   Our new metric has three non-trivial (dimensionless)
parameters, which is one more than the number in the rotating black
hole.\footnote{The rotating black hole has three parameters if one
counts the mass, and the two angular momenta.  But one of these is ``trivial''
in the sense that it can be absorbed into an overall scaling of the
metric.}

   We identified three limiting cases of the new metrics that are of 
particular interest.
{\it Case I} is a limit that gives back the standard rotating black
hole, with two independent rotation parameters.  {\it Case II}
is a limit that gives the original single-rotation black ring, which was
found in \cite{empereal}.  {\it Case III} is a limit giving rise to a
new family of static metrics, with two non-trivial dimensionless parameters.

  Having found the local form of the new metrics, we then studied their
global structure.  For the static metrics obtained in the {\it Case III}
limit, we found that the conditions following from requirement of
no conical singularities imposes periodicity conditions on the azimuthal
coordinates which imply that the horizon has the topology of the lens space
$L(n;m)$, where $m$ and $n$ are positive integers satisfying $m\ge n+2\ge 3$.
The lens space $L(n;m)$ is defined as a factoring $S^3$ by a certain
freely-acting discrete subgroup $\Gamma(n;m)$ of the $SO(4)$ isometry group.
The black hole horizon is an inhomogeneous distortion of the ``round''
lens space.
By contrast, asymptotically at infinity the spacetime approaches
(Minkowski)$_5/\Gamma(m;n)$.  This means that the spatial sections at
large radius are lens spaces $L(m;n)$.  We calculated all the conserved
charges and thermodynamic quantities for these lens-space black holes,
and showed that the first law of thermodynamics is satisfied.  Our
solutions demonstrate that black holes with (Minkowski)$_5/\Gamma(m;n)$
asymptotic structure and a given mass are not unique.

   We then generalised the static metrics to charged solutions within
five-dimensional $N=2$ supergravity coupled to two vector multiplets.
Equivalently, they can be viewed as solutions of five-dimensional 
$N=8$ supergravity, with three independent charges for the $U(1)^3$ gauge
fields in the Cartan subalgebra of $SO(6)$.  Again we found the same
``slumping'' feature exhibited by the uncharged black holes, with 
$L(n;m)$ horizon topology and $L(m;n)$ spatial sections at infinity.

  Finally, we investigated the global structure of the general new
solutions with three non-trivial dimensionless parameters.  For these,
we find that (except for limiting cases that reduce to the previous
discussion) the avoidance of conical singularities now requires that
the time coordinate also be identified periodically.  This is reminiscent
of the situation in the Taub-NUT metrics in four dimensions.  In fact, one
can take the view that the general new solutions we have found are 
the natural five-dimensional analogue of the four-dimensional rotating
Taub-NUT metrics.  (The general construction of higher-dimensional
rotating Taub-NUT metrics in \cite{chenlupope1,chenlupope2} gave only
a trivial ``NUT parameter'' in the special case of five dimensions.)

\section*{Acknowledgements}

   We are grateful to Malcolm Perry for discussions.  Research
supported in part by DOE grant DE-FG03-95ER40917.

\appendix

\section{Rotating Black Ring}\label{ringsec}

   The black ring solution (\ref{ring1}) and (\ref{ringG}), which 
we obtained as a limit of our general metrics (\ref{plebdem51}), 
was previously
obtained in \cite{empereal}.  However, the metric is somewhat simpler in the
coordinates and parameterisation used in (\ref{ring1}).  It is useful,
for completeness, to present a summary of the global structure and the 
thermodynamic quantities in the formalism we are using in this paper.

    We begin by reparemeterising the constants in such a way that $G(\xi)$
in (\ref{ringG}) becomes
\be
G(\xi)=-\mu^2 (\xi-\xi_1) (\xi-\xi_2) (x-\xi_3)\,.
\ee
We can rescale $\mu$ purely by means of scalings of the 
coordinates and parameters $\xi_i$, without needing to rescale the metric.
Since, in this sense, $\mu$ is a trivial parameter, we may set $\mu=1$
without loss of generality.

   To describe the ring, we should take the coordinate $x$ to lie 
in the range $\xi_1\le x\le \xi_2<0$. The coordinate $y$ lies in the 
range $\xi_2 \le y\le \infty$.
The asymptotic region at infinity is located at $x=\xi_2 = y$, and
the horizon is at $y=\xi_3 >0$.   The boundary of the ``ergo sphere'' 
lies at $y=0$.  There is a curvature singularity at $y=\infty$, which  is 
hidden behind the horizon at $y=\xi_3$.  There would also be a curvature
singularity at $1- xy=0$ outside the horizon, which can be avoided by
taking $\xi_1\xi_2 <1$ and $\xi_1 \xi_3 >-1$.   For later convenience,
we introduce two positive parameters $(\eta_1,\eta_2)$,
with 
\be 
\xi_1=-\eta_1^2\,, \qquad \xi_2=-\eta_2^2\,.
\ee

   Having addressed the question of power-law curvature singularities,
we must now examine the possible conical singularities at locations
where the metric degenerates.  First, we note that
the Killing vector $\del/\del\phi$ is degenerate at both $x=\xi_1=-\eta_1^2
$ and $x=\xi_2=-\eta_2^2$.  Normalising to unit Euclidean surface gravity, 
the degenerate Killing vectors at each of $x=\xi_1$ and $x=\xi_2$ are,
respectively,
\be
\ell_1 = \fft{\eta_1}{(\eta_1^2-\eta_2^2)(\xi_3+\eta_1^2)}\, 
  \fft{\del}{\del\phi}\,,\qquad
\ell_2 = \fft{\eta_2}{(\eta_1^2-\eta_2^2)(\xi_3+\eta_2^2)}\,
  \fft{\del}{\del\phi}\,.
\ee
Since these must each generate $2\pi$ rotations, it follows that the two
prefactors must be equal, and hence we must require
\be
\xi_3=\eta_1 \eta_2\,.
\ee

   There is also a degenerate spacelike Killing vector $\ell_3$ at $y=\xi_2$.
It is convenient to make a coordinate transformation to remove a $\del/\del t$
component from this Killing vector.  This is achieved by sending
\be
t\rightarrow \fft{\sqrt{\eta_2}}{\eta_1^{3/2}} t +
\fft{\eta_1^{3/2}}{\sqrt{\eta_2}} \psi\,.
\ee
(We have also scaled the new $t$ variable to give it the canonical 
normalisation for time at infinity.)
Furthermore, we shall define rescaled azimuthal angles $\phi_1$ and 
$\phi_2$, given by
\be
\phi_1 = (\eta_1-\eta_2)(\eta_1+\eta_2)^2\, \psi\,,\qquad
\phi_2= (\eta_1-\eta_2)(\eta_1+\eta_2)^2\, \phi\,.
\ee
In terms of $\phi_1$ and $\phi_2$, the three spacelike Killing vectors
discussed above, which degenerate at $x=\xi_1$, $x=\xi_2$, and $y=\xi_3$,
are given, when normalised to unit Euclidean surface gravity, by
\be
\ell_1=\fft{\del}{\del\phi_2}\,,\qquad \ell_2= \fft{\del}{\del\phi_2}\,,
\qquad \ell_3=\fft{\del}{\del\phi_1}\,.
\ee
Thus we see that $\phi_1$ and $\phi_2$ should both have period $2\pi$.

   Note that $g_{\phi_1\phi_1}$ and $g_{\phi_2\phi_2}$ never become
negative outside the horizon.

       The region near asymptotic infinity can be seen more clearly by
introducing coordinates $r$ and $\theta$, defined in terms of $x$ and $y$ by 
\be
\fft{\sqrt{\xi_2-x}}{y-x}= \fft{(\eta_1+\eta_2)\sqrt{\eta_1-\eta_2} }{
              \eta_2}\, r\, \cos\theta\,,\qquad
\fft{\sqrt{y-\xi_2}}{y-x}= \fft{(\eta_1+\eta_2)\sqrt{\eta_1-\eta_2} }{
              \eta_2}\, r\, \sin\theta\,.
\ee
The metric in the asymptotic region then takes the simple form
\be
ds^2\rightarrow -dt^2 + dr^2 + r^2 (d\theta^2 + \sin^2\theta d\phi_1^2 +
\cos^2\theta d\phi_2^2)\,.
\ee
The ADM mass and the angular momentum can then be straightforwardly
obtained by means of Komar integrals, and are given by
\be
M=\fft{3\pi}{8 \eta_2 (\eta_1^2-\eta_2^2)(\eta_1+\eta_2)}\,,\qquad
J_{\phi_1}=\fft{\pi \eta_1^{3/2}}{4\eta_2^{3/2}(\eta_1^2-\eta_2^2)^2
(\eta_1+\eta_2)^2}\,,\qquad J_{\phi_2}=0\,.
\ee

     The horizon is located at $y=\xi_3$.  The geometry is
a product of $S^2$, with coordinates $(x, \phi_2)$,
and $S^1$, with coordinate $\phi_1$.  It is straightforward
to evaluate the temperature, entropy and angular velocity, leading to
\be
T=\fft{\eta_2(\eta_1 + \eta_2)}{2\pi}\,,\qquad
S=\fft{\pi^2}{2\eta_2 (\eta_1^2-\eta_2^2) (\eta_1+\eta_2)^3}\,,\qquad
\Omega_{\phi_1}= \fft{\sqrt{\eta_2} (\eta_1^2-\eta_2^2)}{\sqrt{\eta_1}}\,.
\ee

  These quantities satisfy the expected thermodynamic relations
\be
dM=TdS + \Omega_{\phi_1} dJ_{\phi_1}\,,\qquad
M=\ft32 (TS + \Omega_{\phi_1} J_{\phi_1})\,.
\ee

\section{Kaluza-Klein Monopole}\label{KKmonosec}

   In this appendix, we examine the static black hole metrics
discussed in section \ref{staticsec} in the limit where $\xi_1=\xi_2$.
Since $x$ in lies in the interval $\xi_1 \le x \le \xi_2$, it
follows that we need to blow up the interval when we take such a limit.
This can be achieved by means of the coordinate and parameter 
redefinitions
\be
\xi_1=\xi_0-\epsilon\,,\qquad \xi_2=\xi_0+\epsilon\,,\qquad
x=\xi_0 + \epsilon \cos\theta\,,\qquad
\phi_2\rightarrow \fft{1}{\epsilon}\phi_2\,,
\ee
in which $\epsilon$ is then sent to zero.  We then introduce a new radial
coordinate $r$ in place of $y$, and also define new azimuthal coordinates
$\phi$ and $\psi$, according to
\be
y=\xi_0 f\,,\qquad
\phi_1=\ft12 (\phi+ \psi)\,,\qquad
\phi_2=\ft12 (\phi- \psi)\,,
\ee
where
\be
f=1 - \fft{\sqrt{1-\xi_0^2}}{2\mu\xi_0\, r}\,.
\ee
The resulting metric then takes the form
\be
ds^2=-f\, dt^2 + \fft{1-\xi_0^2 f}{(1-\xi_0^2)} \Big(\fft{dr^2}{f} +
r^2 (d\theta^2 + \sin^2\theta d\phi^2)\Big) + \fft{1}{4\mu^2 (1-\xi_0^2 f)}
(d\psi + \cos\theta d\phi)^2\,.
\ee
This can be recognised as the metric of the Kaluza-Klein monopole in $D=5$.

\section{A General Class of Higher-Dimensional Local Metrics}

   Here, we record some results for a class of Ricci-flat metrics
in arbitrary dimensions $D\ge4$, which we obtained while searching for 
generalisations of our five-dimensional construction.  We find that
\bea
ds&=& f\Big(\fft{x^2-y^2}{X} dx^2 + \fft{y^2-x^2}{Y} dy^2 +
     \fft{X}{x^2-y^2} (d\phi + y^2 d\psi)^2 +
      \fft{Y}{y^2-x^2} (d\phi + x^2 d\psi)^2\Big)\nn\\
&&+(xy)^\gamma dx^\mu dx_\mu
\eea
is Ricci-flat, 
where
\bea
f &=& \fft{t^{c-1- \ft12 (D-4)\gamma}}{(\alpha + (xy)^c)^2}\,,\qquad
c=\ft12 \sqrt{4 + (D-2)(D-4)\gamma^2}\,,\nn\\
X &=& a_0 x^2 (1 + a_1 x^c - a_2 \alpha x^{-c})\,,\qquad
Y= a_0 y^2 (1 + a_2 y^c - a_1 \alpha y^{-c})\,,
\eea
and $a_0$, $a_1$, $a_2$, $\alpha$ and 
$\gamma$ are constants.  There do not appear to be any 
new and non-trivial regular examples contained within these metrics.

\end{document}